\documentstyle[12pt]{article}

\begin{document}
\begin{flushright}
\baselineskip=12pt
CTP-TAMU-23/97\\
DOE/ER/40717--43\\
ACT-08/97\\
\tt hep-ph/9704439
\end{flushright}

\begin{center}
\vglue 1.5cm
{\Large\bf New $(g-2)_{\mu}$ constraints on light-gravitino models
\\}
\vglue 2.0cm
{\Large Tianjun Li$^{1,2}$, Jorge L. Lopez$^3$, and D.V. Nanopoulos$^{1,2,4}$}
\vglue 1cm
\begin{flushleft}
$^1$Center for Theoretical Physics, Department of Physics, Texas A\&M
University\\ College Station, TX 77843--4242, USA\\
$^2$Astroparticle Physics Group, Houston Advanced Research Center (HARC)\\
The Mitchell Campus, The Woodlands, TX 77381, USA\\
$^3$ Bonner Nuclear Lab, Department of Physics, Rice University\\ 6100 Main
Street, Houston, TX 77005, USA\\
$^4$Academy of Athens, Chair of Theoretical Physics, Division of Natural
Sciences\\ 28 Panepistimiou Avenue, 10679 Athens, Greece\\
\end{flushleft}
\end{center}

\vglue 1.5cm
\begin{abstract}
We show that the gravitino contribution to $(g-2)_{\mu}$ is finite in many
popular supergravity models, including no-scale supergravity and string
and M-theory models. This contribution is greatly enhanced for very
light gravitino masses, and leads to new upper bounds on the ratio of the
smuon to the gravitino mass.
\end{abstract}

\vspace{0.5cm}
\begin{flushleft}
\baselineskip=12pt
April 1997\\
\end{flushleft}
\newpage
\setcounter{page}{1}
\pagestyle{plain}
\baselineskip=14pt

\section{Introduction}
One of the quantities which are measured with great precision in particle
physics is the anomalous magnetic moment of the muon $(g-2)_{\mu}$.
The new E821 experiment at Brookhaven is expected to improve the precision of
the previous measurement by a factor 20~\cite{VH}. This improvement should
be sensitive enough to detect the electroweak contribution to $(g-2)_{\mu}$,
and should provide a way to probe for new  physics, especially
supersymmetry~\cite{LNW,CGW}. The most recent experimental value of the
anomalous magnetic moment of the muon is~\cite{PDG}:
\begin{equation}
a_{\mu}^{\rm exp} \equiv \left({{g_{\mu}-2}\over 2}\right)^{\rm exp}=
 (11659230 \pm 84) \times 10^{-10} \ ,
\end{equation}
while the latest theoretical prediction within the Standard Model is estimated
to be~\cite{CGW}:
\begin{equation}
a_{\mu}=(11659181 \pm 15) \times 10^{-10}\ ,
\end{equation}
leaving an allowed interval for new physics  contributions, at the 90\% C. L,
\begin{equation}
-90 \times 10^{-10} < \delta a_{\mu} < 190 \times 10^{-10}\ .
\label{eq:range}
\end{equation}

Recently, light-gravitino models have attracted renewed attention because they
may explain the $ee\gamma\gamma + E_{\rm T, miss}$ event reported by the CDF
Collaboration~\cite{Park}, and may have further interesting consequences at
LEP~2~\cite{JLDN,LNZ}. If the gravitino is very light it will contribute
significantly to the anomalous magnetic moment of the muon~\cite{MO, FDA}. In
addition, in supergravity models the gravitino contribution may diverge, unless
certain relations among the  K\"ahler function components are
satisfied~\cite{FDA}. These relations provide new constraints on supergravity
models, as we believe that $(g-2)_{\mu}$ should be finite in any
consistent theory after spontaneous supergravity breaking. We have checked
that $(g-2)_{\mu}$ is indeed finite in various models based on no-scale
supergravity, weakly-coupled heterotic strings, and M-theory-inspired models.
We also present numerical studies of the gravitino contribution to
$(g-2)_{\mu}$, and the upper bound that results on $(m_{\tilde\mu}/m_{\tilde
G})$ when $\delta a_\mu$ is restricted to the range in Eq.~(\ref{eq:range}).

\section{Finite gravitino contributions}
In general, the one-loop gravitino contribution to $(g-2)_{\mu}$ may be
divergent because of the non-renormalizability of supergravity. It is
interesting to explore under what conditions will this contribution be
finite in spontaneously broken supergravity. Using dimensional reduction and
requiring one-loop finiteness independently of the photino mass, del Aguila
obtained the following constraints on the K\"ahler function~\cite{FDA}:
\begin{equation}
\langle G_z^{ij}\rangle= 0\ ,\qquad \langle G_{zz}^{ij}\rangle= 0\ ;
\qquad (i\not=j)
\label{eq:conditions}
\end{equation}
where $i,j$ indicate derivatives with respect to the charged sleptons, and $z$
represents derivatives with respect to hidden sector fields (including the
dilaton and moduli fields). These constraints have been obtained in a field
basis where \cite{FDA}
\begin{equation}
\langle G^m_n\rangle=-{\textstyle{1\over2}}\,\delta^m_n\ ,
\label{eq:aux}
\end{equation}
which assures canonical kinetic terms.

First let us consider standard no-scale supergravity~\cite{no-scale}, where the
gauge kinetic function is $f_{\alpha \beta} = h(z) \delta_{\alpha \beta}$ and
the K\"ahler function for the ${{SU(N,1)}\over {SU(N)\times U(1)}}$ theory is:
\begin{equation}
G= -3\ln(z+z^*+\phi_i\phi_i^*)\ ,
\end{equation}
where $z$ is a hidden-sector (singlet) field and the $\phi_i$ represent
observable sector fields. Without loss of the generality, this K\"ahler
potential satisfies the field basis condition in Eq.~(\ref{eq:aux}) when we
choose $\langle z\rangle ={{\sqrt 6}\over2}$ and $\langle\phi_i\rangle=0$.
This choice also allows $G$ to satisfy the finiteness conditions in
Eq.~(\ref{eq:conditions}).

Next we consider an $E_6\times E_8$ compactified model derivable from the
weakly-coupled heterotic string or M-theory~\cite{Witten85, LLN}.
The K\"ahler potential is:
\begin{equation}
G=-\ln(S + S^*) - 3\ln(T+T^*-\phi_i \phi_i^*) \,
\end{equation}
and the gauge kinetic function is $f_{\alpha\beta} =S\,\delta_{\alpha\beta}$.
This K\"ahler potential satisfies the field basis condition in
Eq.~(\ref{eq:aux}) when we choose $\langle S\rangle={{\sqrt2}\over2}$,
$\langle T\rangle ={{\sqrt 6}\over 2}$, and $\langle\phi\rangle =0$. Clearly
the dilaton field satisfies the finiteness conditions
[Eq.~(\ref{eq:conditions})] for $z=S$. The $T$ field also satisfies these
conditions, as shown in the above no-scale supergravity case.\footnote{In fact,
we note that in string-derived models there are no such
diagrams as Fig. 1(e) in Ref.~\cite{FDA}, because of the absence of the
$\gamma$-$\gamma$-$T$ vertex. Therefore, $T$ only needs to satisfy the
finite K\"ahler potential condition: $\langle G_T^{ij}\rangle =0$.}

Let us also consider string no-scale supergravity~\cite{SNS}, where at present
only the lowest-order K\"ahler function is known:
\begin{equation}
G=-\ln(S+ \bar S) +\sum_{I=1,2,3}K_{(I)} +K_{TS} + \ln|W|^2\,
\end{equation}
with
\begin{equation}
K_{(I)}= -\ln\left[1-\sum_i^{n_I}\alpha_i {\bar \alpha}_i
+{1\over 4}(\sum_i^{n_I}\alpha_i^2)(\sum_i^{n_I} {\bar \alpha}_i^2)\right]\ ,
\end{equation}
where $n_I$ represents the number of untwisted fields in set I, and set-indices
$I=1,2,3$ on the $\alpha_i$ (i.e., $\alpha_i^{(I)}$) are understood. Also,
\begin{eqnarray}
K_{TS}&=& \sum_i^{n_{T1}} \beta_i^{(1)} {\bar \beta}_i^{(1)}
e^{{1\over 2}[K_{(2)} +
K_{(3)}]} +\sum_i^{n_{T2}} \beta_i^{(2)}{\bar \beta}_i^{(2)} e^{{1\over 2}
[K_{(1)} +K_{(3)}]}\nonumber\\&&
+\sum_i^{n_{T3}}\beta_i^{(3)}{\bar \beta}_i^{(3)}
e^{{1\over 2}[K_{(1)}+K_{(2)}]}\,
\end{eqnarray}
where the $\beta_i^{(I)}$ are twisted sector fields that belong to the I-th
set, $n_{T1, T2, T3}$ are the number of these fields, and $K_{(1, 2,3)}$
are given in above equation. We see that the constraints in
Eq.~(\ref{eq:conditions}) are satisfied because there are no mixing terms
$\beta_i^{(I)}{\bar \beta}_j^{(J)}$ where $i\not=j$ or $I\not=J$.

Other string-derived orbifold models have similar
K\"ahler potentials at lowest order~\cite{BIM}, for example:
\begin{equation}
K(S, S^*, T, T^*, C_i, C_i^*) = -\ln(S+S^*) +K_0(T+T^*) +
{\tilde K}^i_j(T, T^*) C_iC^{*j}\,
\end{equation}
For phenomenological reasons related to the absence of the flavor-changing
neutral currents in the low-energy theory, one assumes a diagonal
form for the piece of the K\"ahler potential associated with the matter fields,
${\tilde K}^i_j = {\tilde K}_i \delta^i_j$ ~\cite{BIM}. Therefore, by the same
argument as in string no-scale supergravity, in these orbifold models the
gravitino contribution to $(g-2)_\mu$ is finite. Moreover, if ${\tilde K}^i_j$
is not diagonal, then it should satisfy the condition $\langle {\tilde
K}^i_{Tj}\rangle=0$, in order to keep the gravitino contribution finite.

\section{Numerical analysis}
The gravitino contribution to $(g-2)_{\mu}$ in spontaneously broken
supergravity was calculated in Refs.~\cite{MO,FDA}. In what follows
we explore its magnitude and obtain constraints on the sparticle and
gravitino masses by requiring that it be confined to the experimentally
allowed interval in Eq.~({\ref{eq:range}). The usual supersymmetric
contributions to $(g-2)_{\mu}$ ({\em i.e.}, not including the gravitino)
have been studied in the literature \cite{LNW,CGW} and will not be
addressed here, other than to assume that they themselves satisfy
Eq.~(\ref{eq:range}).

The gravitino contribution depends crucially on the gravitino mass
($m_{\tilde G}$), and is of phenomenological importance only for very light
gravitino masses ($m_{\tilde G}<10^{-4}\,{\rm eV}$). This contribution also
depends
on the smuon ($m_{\tilde\mu_{1,2}}$) and photino ($m_{\tilde\gamma}$) masses,
and is given by \cite{FDA}
\begin{equation}
a_\mu^{\tilde G} \approx {G_N\, m_\mu^2\over6\pi m^2_{\tilde G}}\, \sum_{k=1,2}
\left[ {1\over6}+{m^2_{\tilde\gamma}\over m^2_{\tilde\mu_k}-m^2_{\tilde\gamma}}
\left({m^2_{\tilde\gamma}\over m^2_{\tilde\mu_k}-m^2_{\tilde\gamma}}
\ln{m^2_{\tilde\mu_k}\over m^2_{\tilde\gamma}}-1\right)
+(-1)^k\sin(2\alpha)\,{m_{\tilde\gamma}\over m_\mu}\right] m^2_{\tilde\mu_k}\ .
\label{eq:formula}
\end{equation}
We start our numerical analysis by neglecting the smuon left-right mixing angle
($\alpha$), and assuming degenerate smuon masses, in which case the above
formula reduces to
\begin{equation}
a_\mu^{\tilde G}\approx 8\times10^{-10}
\left({m_{\tilde\mu}\over 100\,{\rm GeV}}\right)^2
\left({10^{-5}\,{\rm eV}\over m_{\tilde G}}\right)^2
f\left(m^2_{\tilde\gamma}\over m^2_{\tilde\mu}\right)\ ,
\label{eq:simple}
\end{equation}
where
\begin{equation}
f(x)={\textstyle{1\over6}}+{x\over 1-x}\left({x\over 1-x}\ln{1\over x}-1\right)
\qquad[f(1)=-{\textstyle{1\over3}}]\ .
\label{eq:f}
\end{equation}
One can then see that for realistic values of the smuon and photino masses
($f(x):{1\over6}\to-{1\over3}$ for $(m_{\tilde\gamma}\ll m_{\tilde\mu})
\to(m_{\tilde\gamma}\approx m_{\tilde\mu})$), and rather light gravitino masses
[$m_{\tilde G}\sim(10^{-4}\to10^{-5})\,{\rm eV}$],
$a_\mu^{\tilde G}$ may be sizeable, observable, and of either sign.

There are further aspects of this result that are worth pointing
out. Note first that even lighter gravitino masses (say $m_{\tilde
G}<10^{-6}\,{\rm eV}$) lead to $a_\mu^{\tilde G}$ values that do not fit
in the allowed interval in Eq.~(\ref{eq:range}). However, such light gravitino
masses are already ruled out experimentally from collider and astrophysical
considerations \cite{LNZ,astro}. Furthermore, the function $f(x)$ has a zero
at $x\approx{1\over4}$, and therefore the (fine-tuned) choice $m_{\tilde\gamma}
\approx {1\over2}m_{\tilde\mu}$ would preclude an absolute lower bound
on $m_{\tilde G}$.

Also of interest is the fact that $a_\mu^{\tilde G}$ grows as
$m^2_{\tilde\mu}$, as a result of the non-renormalizability of supergravity.
Indeed, for a fixed value of $m_{\tilde\gamma}\ll m_{\tilde\mu}$, $f(x)\to1/6$
and $a_\mu^{\tilde G}<190\times10^{-10}$ entails
\begin{equation}
\left({m_{\tilde\mu}\over 100\,{\rm GeV}}\right)
\left({10^{-5}\,{\rm eV}\over m_{\tilde G}}\right) < 12\ .
\end{equation}
This experimental upper bound on the ratio $m_{\tilde\mu}/m_{\tilde G}$ is
new, and will become significantly stricter once the data from E821 begins
to be analyzed. A related bound on the sparticle spectrum in the presence
of very light gravitinos has also been obtained from a theoretical viewpoint
by considering violations of tree-level unitarity \cite{Roy}.

Let us now comment on the effect of $\alpha\not=0$. The smuon left-right mixing
angle, in typical supergravity theories, may be estimated as
$\sin 2\alpha\sim m_\mu\, \mu\tan\beta/m^2_{\tilde\mu}$, where $\mu$ is the
Higgs mixing parameter and $\tan\beta$ the usual ratio of Higgs vacuum
expectation values. This angle plays a crucial role in the usual supersymmetric
contribution to $a_\mu$. In the present case, its effective contribution to the
square bracket in Eq.~(\ref{eq:formula}) goes as $\sim
m_{\tilde\gamma}\,\mu\tan\beta/ m^2_{\tilde\mu}$, which is not negligible,
and may even be enhanced for large values of $\tan\beta$. Note however that
this contribution is suppressed if the smuon masses are close to each other
(because of the $(-1)^k$ factor). Without introducing further unknown
parameters into the calculation, one may conclude that the numerical results
obtained above should remain approximately valid in this more general case.

\section{Conclusions}
We have considered the contribution to the anomalous magnetic moment of the
muon from loops involving light gravitinos. This contribution may be divergent
in generic supergravity models, but we have identified several popular ones
where it is finite, including no-scale supergravity and string and M-theory
derived models. We have also studied the phenomenological aspects of this
contribution and established a new upper bound on the ratio of the smuon to
gravitino masses.

\section*{Acknowledgments}
This work has been partially supported by the World Laboratory. The work of
J.~L. has been supported in part by DOE grant DE-FG05-93-ER-40717 and that of
D.V.N. by DOE grant DE-FG05-91-ER-40633.


\begin{thebibliography}{99}
\bibitem{VH} V. Hughes, in {\em Frontiers of High Energy Spin Physics},
Proceedings of the 10th International Symposium, Nagoya, Japan, 1992, edited by
T. Hasegawa et al. (Universal Academy Press, Tokyo, 1992), pp. 717-722.
\bibitem{LNW}J. L. Lopez, D. V. Nanopoulos, and X. Wang, Phys. Rev.
D{\bf 49} (1994) 366; U. Chattopadhyay and P. Nath, Phys. Rev. D{\bf 53}
(1996) 1648; and references therein.
\bibitem{CGW}M. Carena, G. F. Giudice, C. E. M. Wagner, Phys. Lett.
B{\bf 390} (1997) 234.
\bibitem{PDG}Particle Data Group, L. Montanet, et al.,  Phys. Rev. D{\bf 54}
(1996) 1.
\bibitem{Park} S. Park, in {\em Proceedings of the 10th Topical Workshop on
Proton-Antiproton Collider Physics }, Fermilab, 1995, edited by R. Raja
and J. Yoh (AIP, New York, 1995), p. 62.
\bibitem{JLDN} S. Dimopoulos, M. Dine, S. Raby and S. Thomas, Phys. Rev. Lett.
{\bf 76} (1996) 3494; S. Ambrosanio, G. L. Kane, G. D. Kribs, S. Martin, and
S. Mrenna, Phys. Rev. Lett. {\bf 76} (1996) 3498 and Phys. Rev. D {\bf 54}
(1996) 5395; J. L. Lopez and D. V. Nanopoulos, Mod. Phys. Lett. A{\bf11}
(1996) 2473 and Phys. Rev. D {\bf 55} (1997) 4450.
\bibitem{LNZ} J. L. Lopez, D. V. Nanopoulos, and A. Zichichi, Phys. Rev. Lett.
{\bf 77} (1996) 5168 and Phys. Rev. D{\bf 55} (1997) 5813 and references
therein.
\bibitem{MO}A. Mendez and F. X. Orteu, Nucl. Phys. B{\bf 256} (1985) 181 and
Phys. Lett. B{\bf 163} (1985) 167.
\bibitem{FDA}F. del Aguila, Phys. Lett. B{\bf 160} (1985) 87.
\bibitem{no-scale}J. Ellis, C. Kounnas, and D. V. Nanopoulos, Nucl. Phys. B{\bf
241} (1984) 406 and B{\bf 247}(1984) 373. For a review, see A. Lahanas and D.
V. Nanopoulos, Phys. Rep. {\bf 145} (1987) 1.
\bibitem{Witten85} E. Witten, Phys. Lett. B{\bf 155} (1985) 151.
\bibitem{LLN} T. Li, J. Lopez and D. Nanopoulos, hep-ph/9704247.
\bibitem{SNS}J. L. Lopez and D. V. Nanopoulos, Int. J. Mod. Phys. A{\bf 11}
(1996) 3439.
\bibitem{BIM}A. Brignole, L. E. Iba\~nez and C. Mu\~noz, Nucl. Phys.
B{\bf 422} (1994) 125; A. Brignole, L. E. Iba\~nez, C. Mu\~noz and C.
Scheich, Z. Phys. C{\bf 74} (1997) 157.
\bibitem{astro} J. Ellis, K. Enqvist, and D.~V.~Nanopoulos, Phys. Lett. B
{\bf 151} (1985) 357; T. Gherghetta, Nucl. Phys. B {\bf485} (1997) 25;
J. Grifolds, R. Mohapatra, and A. Riotto, hep-ph/9610458.
\bibitem{Roy} See e.g., T. Bhattacharya and P. Roy, Nucl. Phys. B {\bf 328}
(1989) 469, 481.
\end{thebibliography}
\end{document}